\documentclass[preprint,12pt]{elsarticle}

\usepackage{amssymb}
\usepackage{amsmath}
\usepackage{comment}
\usepackage{hyperref}
\usepackage{listings}
\usepackage{amsthm}
\usepackage{graphicx}
\usepackage{tikz}
\usepackage{float}
\usetikzlibrary{calc,arrows.meta}
\usepackage{booktabs}

\usepackage{lineno}

\newenvironment{acknowledgments}{
  \section*{Acknowledgments}
}{}

\journal{Nuclear Physics B}

\begin{document}

\begin{frontmatter}



\title{Optimising Cylindrical Algebraic Coverings for use in SMT by Solving a Set Covering Problem with Reasons}


%\author{} %% Author name

\author{Abiola T. Babatunde\corref{cor1}\fnref{label2}}
    \ead{babatundea@uni.coventry.ac.uk}
%%\author[1]{Abiola T. Babatunde}[%
%%email=babatundea@uni.coventry.ac.uk,]
%%\address[1]{Coventry University, Coventry, UK}
            
\author{Matthew England\corref{cor2}\fnref{label2}}
    \ead{Matthew.England@coventry.ac.uk}
%\address[1]{}

\author{AmirHosein Sadeghimanesh\corref{cor3}\fnref{label2}}
    \ead{AmirHosein.Sadeghimanesh@coventry.ac.uk}

\affiliation{organization={Coventry University},
            addressline={Priory Street}, 
            city={Coventry},
            postcode={CV1 5FB}, 
            %state={},
            country={United Kingdom}}

%\affiliation{organization={},%Department and Organization

\begin{abstract}
The Conflict-Driven Cylindrical Algebraic Covering (CDCAC) algorithm has proven well-suited for performing the theory validation checks in the Satisfiability Modulo Theories (SMT) paradigm for the theory of non-linear real arithmetic (NRA).  CDCAC repurposes the theory underpinning the classical Cylindrical Algebraic Decomposition (CAD) for SMT: it is implemented in SMT-solvers cvc5 and SMT-RAT, and the computer algebra system Maple.
    
It was previously observed that when using CAD for an SMT theory call the output can be optimised via the solution of a single Set Covering Problem (SCP) instance, minimising the conflict clause.  In this paper we consider the corresponding optimisation for CDCAC: we observe how CDCAC actually offers multiple such optimisations in a single call.  Each time we generalise a covering in one dimension we must label the new cell in the next dimension with those theory constraints which cannot be together satisfied upon it: we seek the smallest subset of constraints which together cover each of the labels from the cells in the covering of the present dimension.  We call this more involved optimisation a ``Set Covering Problem with Reasons'' (SCPR).

To make the SCPR easier, we present a data reduction step — a generalisation of Beasley's reduction for the SCP — and we find that this alone actually solves many of the that SCPR instances arising from the problems in the SMT-LIB. We then suggest an exact solver utilising linear programming to solve the remaining cases efficiently. Integrating these steps into CDCAC would speed up SMT-solver performance for NRA problems.
\end{abstract}


\begin{highlights}
    \item Problem formalisation: We extended the classical Set Covering Problem to a new variety we call the Set Covering Problem with Reasons (SCPR), motivated to model the choice generalising multi‐constraint conflicts in higher‐dimensional CAD cells.
    \item Generalised Beasley reduction: We adapted Beasley's data‐reduction rules~\cite{BEASLEY1987851} to SCPR, enabling elimination of 95.5\% of CDCAC‐derived instances from this purely syntactical reduction technique.
    \item Solver evaluation: For the 4.5\% of instances remaining after reduction, we showed that a linear‐programming (LP) formulation finds optimal reason sets in under 1\,ms per instance, outperforming SAT-solver based and heuristic alternatives.
    \item Practical impact: Integrating this pipeline into CDCAC would yield provably minimal conflict clauses with almost zero overhead, accelerating clause learning in SMT-solvers~\cite{Abraham2021}.
\end{highlights}
 

\end{frontmatter}

\section{Introduction}
\label{intro}

Modern SMT-solvers depend critically on efficient theory engines, especially for Non-linear Real Arithmetic (NRA). Methods based on the theory of Cylindrical Algebraic Decomposition (CAD) are the only implemented complete decision procedures for NRA.  CAD, in its classical formulation \cite{Collins1975}, builds a full decomposition of $\mathbb{R}^n$ which may then be analysed cell by cell: but this is prohibitively expensive in an SMT context where a single problem instance can produce many queries to CAD. The NLSAT/MCSAT framework of Jovanovi\'{c} and de Moura~\cite{JdM12}, and the CDCAC framework of \'{A}brah\'{a}m \emph{et al.}~\cite{Abraham2021} mitigate this by interleaving CDCL-style search with incremental CAD coverings, generalising each conflict to a maximally sized CAD cell to guide future search.  These frameworks preserve CAD’s completeness, while avoiding exhaustive exploration of irrelevant cells.  Nevertheless, they still have scope for significant optimisation.  
This paper is concerned with CDCAC, used within a traditional SMT approach, i.e. CDCL(T) (where the {T} denotes modulo Theories). Here a key driver of performance is the size of the learned conflict clauses: disjunctions of negated constraints that avoid the SAT solver producing the previously proposed solution (and hopefully many others).   If this clause is larger than needed, it rules out less of the search space and thus slows down the SAT-solver and increases its memory use. 

It has been observed that if we use classical CAD for SMT theory calls, one minimises this clause length by casting conflict core extraction as a Set Covering Problem (SCP) over conflicting sample points~\cite{Kremer2020AIB,Kremer2020JSC}. However, CDCAC’s per-cell conflict generalisations often involve combinations of constraints, and so a richer model of the optimisation problem is needed. We identify that in the present paper and formalise it as the ``\emph{Set Covering Problem with Reasons}'' (SCPR): where we cover cells implicitly by first covering the sets of constraints needed to conclude unsatisfiability on them.  

We examine these optimisation problems next in Section \ref{sec:opt}. We then present a new reduction process for the SCPR problems in Section \ref{sec:reduce} (generalising Beasley reduction for the SCP) and a solution process based on linear programming in Section \ref{sec:sol}.  Finally we give experimental results in Section \ref{sec:results} and conclusions in Section \ref{sec:Conc}.  Some of the ideas in Section \ref{sec:opt} and \ref{sec:sol} were previously discussed in the extended abstract \cite{Sadeghimanesh-England-2022}.

\section{Conflict Clause Optimisation for CAD and CDCAC}
\label{sec:opt}

We first explain how set covering optimisation problems emerge from the use of CAD/CDCAC within SMT.  

\subsection{CAD}

Classical CAD algorithms~\cite{Collins1975} proceed in two phases: (i) a \emph{projection} phase that computes from the input a finite set of polynomials whose roots capture the ``interesting'' boundaries where properties of the input set change; and 
(ii) a \emph{lifting} phase that builds CADs incrementally by dimension in reference to those polynomials until we reach the full dimension of the problem.  This is done by working with the polynomials at sample points; using univariate polynomial real root isolation to decompose the next dimension above the point; and then generalising the finding from the point to the cell it came from.  The correctness of such an approach rests in showing that the generalisation from the point is correct due to the information captured in the polynomials by the projection.  
The result of this process is a decomposition of the problem space into cells:  each cell is semi-algebraic, meaning it is described by polynomial constraints, and cylindrically arranged, meaning any two cells projection onto lower dimensions are either identical or disjoint; i.e. the cells stack up in cylinders. 

\begin{figure}[H]
    \centering
    \includegraphics[width=0.4\textwidth]{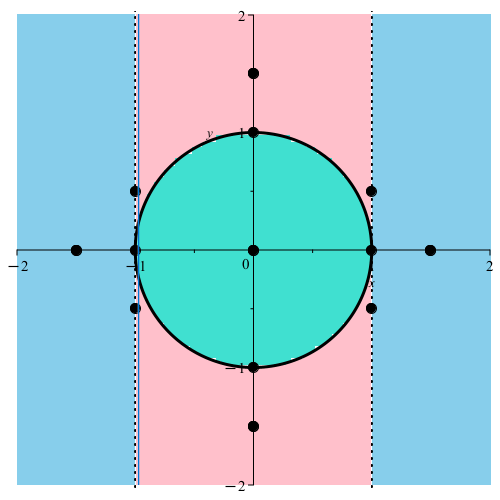}
    \caption{A CAD for the unit circle. It consists of 13 cells: the 13 black discs identify one sample point from each.  The 2 points at the intersection of the circle and the dotted lines are point-cells, the other 6 discs on the circle or dotted lines represent one-dimensional cells for those line and curve segments, with the remaining 5 cells representing two dimensional cells of the plane as identified with the different colours. Note the vertical alignment of sample points, illustrating the organisation of cells into cylinders. \label{fig:EGcad} }
\end{figure}

The original CAD algorithm produced CADs relative to a set of input polynomials so that each polynomial had invariant sign on each cell. For example, consider Figure \ref{fig:EGcad} which illustrates a simple CAD relative to the unit circle.  More advanced algorithms have been produced since which ensure invariance instead for the truth of polynomial formulae~\cite{CH91, Brown1998, EBD20}. 

CAD was originally introduced by Collins in 1975 as a method for quantifier elimination (QF) in real closed fields~\cite{Collins1975}, but now also serves as a decision procedure for (QF)NRA in SMT.  Given a conjunction of polynomial constraints, one can build a CAD sign-invariant for those polynomials and then check if any CAD cell yields a consistent solution.  We need only test one sample point per cell: if any one satisfies the formula, it is $\mathsf{SAT}$; and if none do, the formula is $\mathsf{UNSAT}$. For example, given the SMT problem $\exists\, x, y\;\text{s.t.}\; x^2 + y^2 < 1$ we could build the CAD visualised in Figure \ref{fig:EGcad}, test the 13 sample points and find that satisfying solutions are given only within the green cell (interior of the circle).

In contrast to heuristic approaches like incremental linearisation \cite{CGIRS18c}, CAD's systematic exploration of the real solution space guarantees \emph{completeness}.  However, this comes at the cost of worst-case doubly-exponential complexity \cite{BD07}.  There have been several adaptions of CAD to the SMT context which improve practical efficiency \cite{Kremer2020JSC, JdM12, Abraham2021}, but they all retain this worst case complexity.  Thus many SMT-solvers use a mixture of a complete CAD-based algorithm and a quick heuristic algorithm in practice~\cite{BN24, KRBT22}.

\subsection{CAD conflict clause optimisation}
\label{subsec:CADopt}

In an SMT context, a SAT-solver proposes a solution to the Boolean skeleton of a problem, and then a theory solver checks if this is satisfiable in the theory:  it must produce a conflict clause to explain a finding of unsatisfiability, which is used to modify the following Boolean search. Each such conflict clause is a disjunction of negated constraints (literals) that describes an infeasible combination of constraints. It is desirable for these learned clauses to be as small as possible, i.e. containing fewer constraints, because that will allow for a wider generalisation of the conflict and thus prune more of the search space for the SAT-solver.  

For an NRA theory solver based on CAD this optimisation can be formalised as classical \emph{Set Covering Problem} (SCP).  In general, an SCP is defined by a universe set $U$ and a collection $E$ of subsets of $U$; with the problem to identify the smallest number of sets from $E$ whose elements together cover $U$.  In this context, we want to find the fewest number of constraints such that at least one is not satisfied on each cell of the CAD.  This is equivalent to an SCP considering $U$ as the set of cells in the CAD and $E$ as the collection of subsets: one set for each constraint containing those cells on which the constraint is not satisfiable.

Prior work on this optimisation includes the method of Jaroschek, Dobal and Fontaine~\cite{JDF15}, who cast the minimal conflict–set problem as a $0‑1$ integer linear program (ILP) to be solved via linear or mixed integer programming. Here, a $0‑1$ ILP (also called a binary integer program) is a linear optimisation model with linear constraints in which each decision variable is restricted to $\{0,1\}$, which in this setting indicates whether a constraint is selected ($1$) or not ($0$). Hentze’s thesis introduced a preconditioning step and a hybrid set‑cover algorithm that alternates between a greedy heuristic and exhaustive enumeration depending on problem size~\cite{Hentze2017}. Kremer’s dissertation (2020) also adopts this hybrid approach~\cite{Kremer2020AIB}.  They report that in practice, a simple greedy set‑cover heuristic usually suffices to find a near‑minimal conflict clause quickly, but when the preconditioning step reduces the problem size sufficiently it can be beneficial to replace the greedy step with an exact search that enumerates all combinations of the remaining constraints~\cite{Hentze2017, Kremer2020AIB}. 
We note that the way constraints are formulated can also impact the efficiency of various CAD optimisations (see e.g. \cite{BDEW13}) and thus future CAD calls. 

\subsection{CDCAC}

The Conflict-Driven Cylindrical Algebraic Covering (CDCAC) algorithm \cite{Abraham2021} reformulated CAD to better suite the SMT computational framework.  
In the case of satisfiability for SMT we need only a single satisfying point, and thus rather than building a whole CAD and then analysing each cell, CDCAC constructs and analyses one cell at a time; and rather than building cells and then choosing samples, CDCAC tries a sample point and then if unsatisfiable builds a sign-invariant cylindrical cell around it, generalising the conflict.  This is done using CAD’s projection‐and‐lifting machinery, but we can make savings as we need only locally relevant information \cite{NASBDE24}.  In the case of  unsatisfiability CDCAC must determine the infeasibility of the whole space by constructing a covering of cells.  This is advantageous compared to a decomposition as we can use bigger, and thus fewer, cells and in particular we can sometimes avoid lower dimensional cells which require costly algebraic number computations.  We note that the NLSAT approach of \cite{JdM12} also implicitly produces coverings, but CDCAC stays within the CDCL(T) framework rather than MCSAT.

Whenever CDCAC determines a conjunction of constraints unsatisfiable, the algorithm has implicitly covered the assignment space by $\mathsf{UNSAT}$ CAD cells. However, these cells are not constructed explicitly, as the algorithm instead works dimension by dimension with a generalisation at each step from sample point to interval. If, at a given dimension, a partial solution cannot be extended, then CDCAC generates a covering of the dimension formed of cells within which the partial model cannot be extended for the same reason.  The algorithm projects these findings back to the lower dimension, creating a new unsatisfiable cells there.  The generalisation is built according to the real roots of the projection polynomials, borrowing theory from CAD.   See \cite{Abraham2021} for the full algorithm specification: we illustrate the main ideas with a small worked example.

\subsection{Worked example of CDCAC}
\label{subsec:cdcac_unsat_example}

Consider the three non‑linear constraints over $\mathbb{R}^2$:
\[
  c_1 : y > x + 1,\qquad
  c_2 : y > 1 - x,\qquad
  c_3 : y < 0.
\]
Geometrically, $c_1$ and $c_2$ require $y$ to lie above two intersecting lines, while $c_3$ forces $y$ below the $x$‑axis. There is no $(x,y)\in\mathbb{R}^2$ satisfying all three constraints, so the formula $c_1\wedge c_2\wedge c_3$ is unsatisfiable.

\begin{figure}[H]
    \centering
    \includegraphics[width=0.24\linewidth]{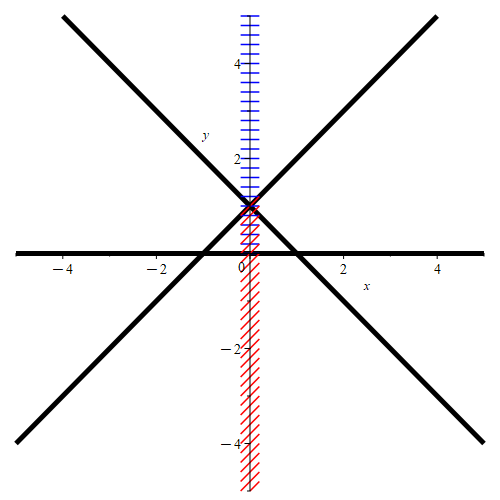}
    \includegraphics[width=0.24\linewidth]{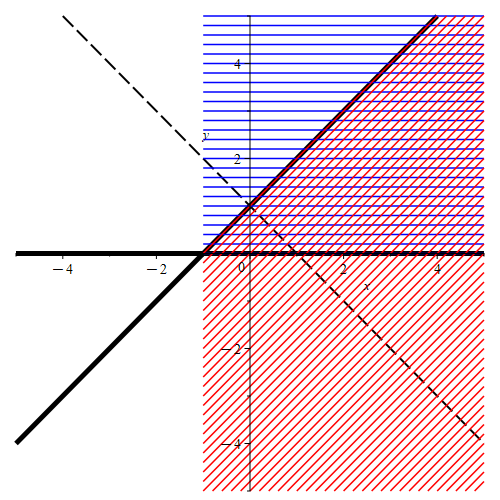}
    \includegraphics[width=0.24\linewidth]{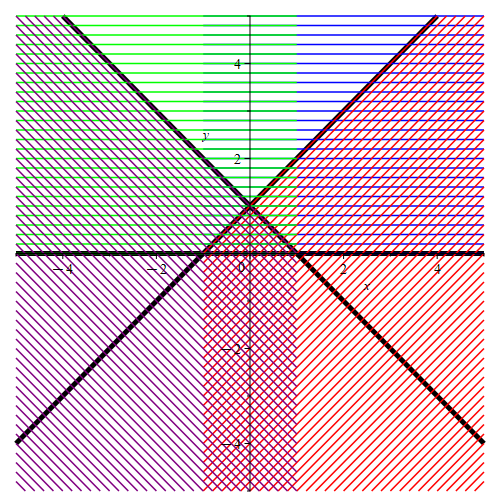}
    \includegraphics[width=0.24\linewidth]{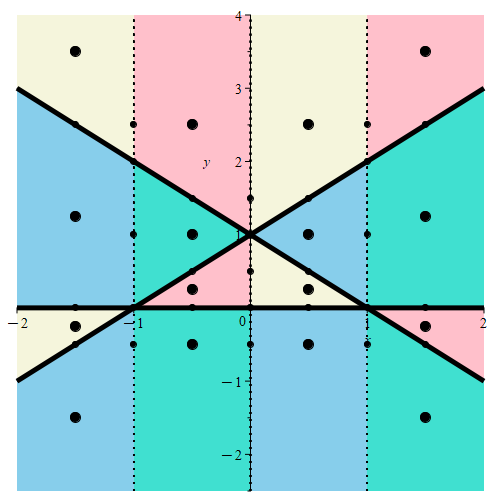}    
    \caption{Following the worked example from left to right: we first analysed at $x=0$, found two unsatisfiable intervals and then generalised the finding to apply for $x \in (-1, \infty)$.  A similar analysis at $x=-3$ completed the covering with 4 cells to prove unsatisfiability of $\{c_1, c_2, c_3\}$.  The right most image shows the full CAD with 43 cells for the three defining polynomials as a comparison. \label{fig:EGCDCAC}}
\end{figure}

Assuming variable ordering $y \succ x$, CDCAC would start by exploring the $x$-dimension. As there are no constraints solely in $x$ it has a free choice and selects $x=0$. It then explores the $y$-dimension to see if this partial sample can be extended.  The constraints now simplify to $y>1$ and $y<0$. Thus we form two unsatisfiable cells that together cover the whole dimension:
\begin{description}
\item[$I_{x=0,1}=(-\infty,1)$] unsatisfiable because of $c_1$ or $c_2$; and
\item[$I_{x=0,2}=(0, \infty)$] unsatisfiable because of $c_3$.
\end{description}
We now need to generalise the finding to rule out an interval in the $x$-dimension.  This involves taking the CAD projection of the constraints defining the cells.  In this case, we have a redundancy in that the first cell is unsatisfiable for both $c_1$ and $c_2$, so we may use only one of them in generalisation: we choose $c_1$.  The projection then produces $x+1$ whose sole root is $x=-1$ which bounds the generalisation.  Thus the conclusions at $x=0$ are generalised to interval $I_{1}=(-1, \infty)$ in the $x$-dimension.  This is visualised in the left two images of Figure \ref{fig:EGCDCAC}.

The algorithm must now proceed to sample $x$ outside this interval.  Choosing $x=-3$ simplifies the constraints to $y>-2, y>4$, and $y<0$ giving us intervals:
\begin{description}
\item[$I_{x=-3,1}=(-\infty,-2)$] unsatisfiable because of $c_1$;
\item[$I_{x=-3,2}=(-\infty,4)$] unsatisfiable because of $c_2$;
\item[$I_{x=-3,3}=(0, \infty)$] unsatisfiable because of $c_3$.
\end{description}
We can actually form the covering using only the latter two cells, allowing us to ignore $c_1$ this time.  The generalisation will produce polynomial $x-1$ whose sole root at $x=+1$ bounds the generalisation to interval $I_{2}=(-\infty,1)$.  We now have two cells covering the whole $x$-dimension and so conclude unsatisfiability of the whole problem.

Note that we have implicitly covered the plane with just four cells.  This compares with a 43 cell CAD for the three constraint polynomials: the comparison is visualised in the right two images of Figure \ref{fig:EGCDCAC}.

\subsection{CDCAC conflict clause optimisation}
\label{sec:CDCACopt}

If CDCAC concludes $\mathsf{UNSAT}$ then we must construct a conflict clause to give back to the SAT-solver, as with CAD.  In the simple worked example above we saw that all three constraints were used in the analysis and so the conflict clause is simply $c_1 \land c_2 \land c_3$.  In general, every unsatisfiable interval constructed must be tagged with the constraints allowing for that unsatisfiable conclusion.  

Observe now the existence of different optimisation problems.  First, consider the case of intervals covering the ``top'' level in the analysis: that is $y$ in the worked example; and in general when we have a partial sample of dimension one lower than the problem. For each such interval generated we can find at least one, and possibly more, of the original constraints that are unsatisfied.  When generalising such a covering we can optimise by minimising the number of constraints in use.  We did this in the worked example when we used only $c_1$ and $c_3$ to generalise around $x=0$ (ignoring $c_2$ as the interval $I_{x=0,1}$ was already concluded unsatisfiable by $c_1$). This optimisation takes the form of an SCP, similar to the case for CAD described in Section \ref{subsec:CADopt}, with $U$ the set of intervals and $E$ a collection of subsets, one for each constraint, identifying the intervals where it is not satisfied.

However, the optimisation at other levels of analysis is different.  The worked example contained only two layers but suppose there were a third variable $w$, the variable ordering were $y \succ x \succ w$, and the analysis in the worked example was done at a partial sample for $w$.  Then the next step of the algorithm would be to generalise to an unsatisfiable interval in the $w$ dimension.  In this case we have two intervals forming the covering:
\begin{description}
\item[$I_{1}=(-1,\infty)$] unsatisfiable for $c_1$ and $c_2$ together; and
\item[$I_{2}=(-\infty,1)$] unsatisfiable for $c_2$ and $c_3$ together.
\end{description}
It would not be correct to treat this as an SCP and take just $c_2$: because although $c_2$ was used to conclude unsatisfiability in both intervals, it could only do so \emph{in combination} with other constraints.  In this example, there is no optimisation to be made: all three constraints are needed to conclude unsatisfiability.  But how would we tackle such an optimisation problem in general?  We formalise this problem next and then consider methods to tackle the problem in the remainder of the paper.

\subsection{The Set Covering Problem with Reasons (SCPR)}

To handle this optimisation problem, we define a new generalisation of the Set Covering Problem, which we call the \emph{Set Covering Problem with Reasons}.

Suppose we have a universal set, $U = \{\alpha_1,\alpha_2,\dots,\alpha_n\}$, $n \in \mathbb{N} $; a set of reasons, $R = \{\rho_1,\rho_2,\dots,\rho_r\}$, $r \in \mathbb{N} $; and a set of pairs of subsets, $E \subseteq \mathcal{P}(U) \times \mathcal{P}(R)$.  I.e. 
$ 
E = \{(A_1,R_1),\dots, (A_m,R_m)\} 
$
where $A_{i} \in$ $\mathcal{P} (U)$ are called covering sets, and $R_{j} \in$ $\mathcal{P} (R)$ are called reason sets. 
    
Then we define the \emph{Set Covering Problem with Reasons (SCPR)} as the problem of finding the minimal subset of reasons such that the covering sets $A_i$ whose accompanying reason sets $R_i$ are covered, themselves cover the entire universal set $U$.  I.e. $S \subset R$ is the minimal subset of $R$ such that
\[
\bigcup_{i \text{ such that } R_i \subset S} A_i = U.
\] 

\subsection{SCPR for CDCAC Optimisation}

This SCPR models the second of the optimisation problems outlined for CDCAC in Section \ref{sec:CDCACopt}: with $U$ being the set of intervals forming the covering, $R$ being the set of constraints, and $A_i$ being the collection of cells for which the combination of reasons in $R_i$ determines unsatisfiability.  To minimise conflict clause length we must minimise the number of reasons such that in combination they cover enough reason sets $R_i$ so that the corresponding interval sets $A_i$ together cover all intervals $U$.

\newpage 

\section{A Generalisation of Beasley Reduction to SCPR}
\label{sec:reduce}

It is well known that instances of optimisation problems derived from applications may be open to simplification before being tackled by full solvers.  The key principles of such a reduction are:
 \begin{enumerate}
     \item any solution to the reduced problem instance can be easily extended to a solution of the original problem instance; 
     \item if the reduced problem does not have a solution, then neither does the original problem instance; and
     \item the cost of reduction should be small compared to the cost of a full solver.
 \end{enumerate}
For the classical SCP there is a well known technique called Beasley reduction  \cite{BEASLEY1987851} which we recap first, before generalising to the case of SCPR.  

\subsection{Beasley reduction for SCP}

Given an SCP instance $(U,E)$, where $U$ is the universe and $E\subseteq 2^U$ the family of subsets, Beasley's process applies two rules iteratively.

\begin{enumerate}
  \item \textbf{Unique Coverage:} If an element $u\in U$ appears in exactly one subset $S\in E$, then $S$ must belong to every optimal cover.  Add $S$ to the partial solution $C^*$, remove $S$ from $E$, and delete all elements of $S$ from $U$.
  \item \textbf{Dominance:} If $S_i,S_j\in E$ satisfy $S_j\subseteq S_i$, then $S_i$ is redundant and can be removed from consideration of inclusion in the final solution (since $S_j$ covers at least as much as $S_i$ with no extra cost).
\end{enumerate}

After no more reductions apply, the algorithm outputs the reduced instance $(U',E')$ and the fixed subsets $C^*$.  Solving $(U',E')$ optimally gives $C'$, and $C'\cup C^*$ is optimal for $(U,E)$ \cite{BEASLEY1987851}.

\subsection{Generalised Beasley reduction for SCPR}

SCPR problem instances are triples $(U,R,E)$, where $U$ is a set of conflict regions, $R$ a set of reasons (constraints), and $E$ is the set of pairs of subsets: $E=\{(A_i,R_i)\mid A_i\subseteq U,\;R_i\subseteq R\}$. We seek a minimal $S\subseteq R$ so that every region in $U$ is explained by at least one $R_i\subseteq S$.  To reduce SCPR we may use the following two rules, which generalise Beasley's rules for the SCP above.

\begin{enumerate}
  \item \textbf{Unique Coverage in SCPR:} If a region $u\in U$ appears in exactly one pair $(A_i,R_i)\in E$ (i.e.\ $u\in A_i$ and in no other $A_j$), then $R_i$ is necessary. Add $R_i$ to the fixed reasons $S^*$, remove all $u'\in A_i$ from $U$, and discard any pairs whose $A_j$ becomes empty.
  \item \textbf{Dominated Pairs in SCPR:} If $(A_j,R_j),(A_i,R_i)\in E$ satisfy $A_j\subseteq A_i$ and $R_i\subseteq R_j$, then $(A_j,R_j)$ is redundant as we can cover that and more with fewer reasons, and so $(A_j,R_j)$ can be removed.
\end{enumerate}

Iterating these rules yields reduced $(U',R',E')$ and fixed reasons $S^*$. Any optimal solution $S'$ for the reduced instance extends to $S'\cup S^*$, which is optimal for the original SCPR. 

We see in the experimental results later that  applying these rules to SCPR instances generated by using CDCAC on SMT-LIB benchmarks actually solves a great many cases outright and significantly shrinks the remainder,  making the subsequent optimisation trivial in many cases.

\section{Solvers for the SCPR}
\label{sec:sol}

Solving the SCPR, like any combinatorial optimisation problem, may demand balance between optimality and speed of solution leading us to consider both exact and heuristic solution methods. In the following, we present a Linear Programming (LP) formulation that guarantees optimal solutions. We also experimented with other exact and heuristic methods: SAT‐based encodings, Partial MaxSAT, and inexact metaheuristics methods. Our experiments on CDCAC derived instances show that the LP‐based solver combined with the generalised Beasley reduction described above is suitable for the problems currently tackled by SMT solvers.

\subsection{Linear programming solution}

To obtain exact, optimal solutions for the SCPR, we encode the problem as a binary linear program. Let each reason \(r_i\in R\) correspond to a binary variable \(y_i\in\{0,1\}\), where \(y_i=1\) if and only if \(r_i\) is selected in the solution. The objective is then
\[
\min\;\sum_{i=1}^{|R|} y_i,
\]
i.e. to minimise the total number of reasons selected.  What constraints are we subject to?

In the simpler SCP case we need to ensure each cell is covered by at least one of a few possible reasons, leading to one constraint for each cell, in which the sum of relevant $y_i$ is forced $\geq 1$, completing the linear program.

However, for the more complex SCPR case we need to ensure a combination of reasons are selected: a reason set \(R_k=\{r_{i_1},\dots,r_{i_t}\}\) leads us to consider a constraint $y_{i_1}\cdots y_{i_t}\geq 1$.  We then have a non-linear constraint for each cell (sums of such monomial if there are different ways to conclude unsatisfiability).  We can restore linearity by introducing auxiliary binary variables for each such monomial. However, we then need to add additional corresponding constraints to ensure consistent vanishing behaviour, i.e. that the new auxiliary binary variable vanishes if any of its factors do, and that if all factors are $1$ so is the auxiliary variable.  To replace the monomial $y_{i_1}\cdots y_{i_t}\geq 1$ with $u_k$ we need to impose 
\[
  u_k \,\le\, y_{i_1},\quad u_k \,\le\, y_{i_2}, \dots, y_{i_t}, \quad
  y_{i_1}+y_{i_2}+\cdots+y_{i_t}-u_k \,\le\, t-1,
\]
so that \(u_k=1\) if and only if all \(y_{i_j}\) are \(1\).

We have now recovered a linear program, although at the expense of greatly increasing both variables and constraints.
We solve the resulting linear program using the \texttt{scipy} optimisation library in Python. 

In our experiments detailed later on CDCAC‐derived instances (after Beasley reduction), the LP approach found proven optimal solutions for all test problems and ran in under one minute on the largest benchmarks, outperforming other exact methods. Combining LP with preprocessing thus yields a practical, high‐performance exact solver for the SCPR.

\subsection{Boolean programming solutions}

We also explored alternative exact solutions to the SCPR using Boolean encodings.  We can move easily to such an encoding from the non-linear optimisation problem above by replacing each binary variable \(y_i\) with a Boolean variable, multiplications with conjunctions, and additions with disjunctions. 
The Boolean optimisation problem is then to minimise the number of \(y_i\) assigned \texttt{True}, subject to the constraints. 

\subsubsection*{Iterative SAT-solver calls}

We first used a SAT-solver by employed an iterative deepening strategy: at each step we impose a cardinality constraint that exactly \(k\) variables are \texttt{True}, generate a formula with all combinations of size \(k\), and then query a SAT-solver (we use Z3) as to whether there is a satisfying assignment of this subject to the constraints. We start with \(k=1\) and increment until a satisfying instance is found, which we can then conclude to have the minimal number of variables true.

Our experiments later show this approach produces exact solutions and performed competitively on small instances, though it was generally far slower than the LP formulation on larger SCPR benchmarks.

\subsubsection*{Partial MaxSAT}

We also encoded the SCPR as a Partial MaxSAT instance: where constraints are marked soft or hard and we seek to maximise how many soft clauses we satisfy while ensuring hard clauses are all satisfied~\cite{fu2006solving}.  For each region \(A_j\) the coverage clause \(\bigvee_{r_i\in R_j} y_i\) is treated as a hard clause, while each unit clause \(\neg y_i\) is a soft clause of weight 1. A MaxSAT solver thus minimising the number of selected reasons, while ensuring we have a covering. We use the implementation in the \texttt{Optimisation} solver in {Z3} which uses the \emph{MAXSAT resolution (MAXRes)} algorithm~\cite{BONET2007606}.  In our experiments later, this method achieved near‐optimal results and competitive runtimes for moderate instance sizes, but occasionally returned solutions one reason larger than the optimum.

\subsection{Metaheuristic solutions}

It is common that for large instances of combinatorial optimisation problems an exact solution is infeasible for exact solution, and thus the problem is tackled by heuristic methods instead.  We adapted some common metaheuristic ideas for solving the SCPR.

In all cases a candidate solution is represented by a bitstring \((y_1,\dots,y_{|R|})\), with \(y_i=1\) indicating that reason \(r_i\) is selected. A fitness function counts the number of reasons selected and adds a large penalty if the chosen reasons fail to cover all regions. 

\subsubsection{Optimisation problem for heuristics}

To apply heuristic and metaheuristic algorithms to the SCPR, we recast the problem into a maximisation problem:
\begin{equation}\label{eq:SCPR_max}
\max \left(r - \sum_{i=1}^r y_i\right) \quad \text{s.t.} \quad \sum_{i=1}^n \chi_{C_i} = n.
\end{equation}
Here $r$ is the number of reasons, $n$ is the number of constraints which is at most the number of elements of $U$, $\chi_{C_i}$ is the binary indicator for whether the constraint $C_i$ is satisfied and $C_i$ is a constraint to indicate that universal set member $\alpha_i$ is covered.

We aim to merge the constraints and objective into a single fitness function, which we define as:
\begin{equation}\label{eq:new_target}
\phi = r - \sum_{i=1}^r y_i + a \sum_{i=1}^n \chi_{C_i},
\end{equation}
where \( a \) is a constant that weights the constraint satisfaction term. We choose $a=r+1$ and use Table~\ref{tab:maximisation_logic} to explain the rationale of that choice: we see that the cases where constraints are not satisfied give worse fitness function values.

\begin{table}[htbp!]
    \centering
    \caption{Illustration of $\phi$ values over different feasible and infeasible configurations.}
    \label{tab:maximisation_logic}
    \begin{tabular}{|l|l|}
        \hline
        \textbf{Search Space Configuration} & \textbf{Target Value $\phi$} \\
        \hline
        $\chi_{C_1} = \cdots = \chi_{C_l} = 1,\; y_1 = \cdots = y_r = 0$ & $r + l(r+1)$ \\
        $\chi_{C_1} = \cdots = \chi_{C_l} = 1,\; \text{exactly one } y_i = 1$ & $r - 1 + l(r+1)$ \\
        $\vdots$ & $\vdots$ \\
        $\chi_{C_1} = \cdots = \chi_{C_l} = 1,\; \text{all } y_i = 1$ & $l(r+1)$ \\
        \hline
        At least one $\chi_{C_i} = 0$ & $\leq l(r+1) - 1$ \\
        \hline
    \end{tabular}
\end{table}

\subsubsection{Heuristic algorithms implemented}

Heuristics were implemented for the SCPR based on the following metaheuristic strategies which had been previously specialised to the SCP.
\begin{itemize}
  \item \textbf{Greedy:} Repeatedly add the reason whose selection covers the largest number of currently uncovered regions until all regions are covered, following \cite{chvatal1979greedy} for the SCP.
  \item \textbf{Genetic Algorithm (GA):} Maintain a population of bitstrings corresponding to reasons selected; select parents proportionally to fitness, and produce offspring via crossover and mutation. The best individual after a fixed number of generations approximates the minimal reason set, following \cite{BEASLEY1996392} for the SCP.
  \item \textbf{Simulated Annealing (SA):} Starting from a random bitstring, at each step flip one bit to produce a neighbour and accept it based on a probability that is gradually decreasing (the temperature). This enables escapes from local minima while converging towards low–cost solutions~\cite{Nikolaev2010}.
  \item \textbf{Particle Swarm Optimisation (PSO):} Represent each particle as a vector of continuous values in \([0,1]^{|R|}\); update its velocity based on personal and global best positions. At evaluation time threshold each coordinate to obtain a bitstring. The swarm converges towards promising regions of the search space; following \cite{balaji2016new} for the SCP.
  \item \textbf{Multi‐Level Score Element‐State Configuration Checking \\(MLSES‐CC):} A local search algorithm that assigns scores to each bit based on how many new regions it covers and flips bits while avoiding recently visited configurations. It continues until a stopping criterion or a coverage threshold is met; following \cite{wang2021improved} for the SCP.
\end{itemize}

Hyperparameters (population size, crossover and mutation rates for GA; cooling schedule for SA; inertia weight and acceleration coefficients for PSO; neighbourhood size and restart threshold for MLSES‐CC) were tuned via a 100–trial tree‐of‐Parzen search~\cite{NIPS2011_86e8f7ab}. 

Our experiments later on CDCAC–derived SCPR benchmarks, show these heuristics achieving $90-95$\% of the optimal reason–set size after Beasley reduction, with SA frequently the fastest. 

\section{Experiments}
\label{sec:results}

To validate our SCPR pipeline on CDCAC‐derived instances, we extracted conflict events when solving problem instances from QF\_NRA benchmarks directory of the SMT‐LIB~\cite{barrett2016smt} with the CDCAC implementation described in \cite{Sadeghimanesh-England-2022}.  The SMT-LIB benchmarks are produced by the SMT community and represent a large and representative dataset for the current applications of SMT-solvers (see Table~\ref{tab:benchmarks} for the scope of problem domains).

\subsection{Data acquisition}

We began with the SMT‐LIB 2024~\cite{barrett2016smt} non‐incremental QF\_NRA suite, sampling one representative file from each of the sixteen top‐level directories (Table~\ref{tab:benchmarks}). Using a Maple 2023 CDCAC implementation with SCPPack~\cite{Sadeghimanesh-England-2022}, we did the following.
\begin{enumerate}
  \item Parsed and executed CDCAC on each benchmark, logging every conflict as a pair \((A, R_A)\), where \(A\) is an unsatisfiable CAD cell (set of sample‐point identifiers) and \(R_A\) the minimal constraints falsified on \(A\).
  \item Excluded four files that failed to parse or timed out, yielding 12 benchmarks.
  \item Post‐processed logs with SCPPack to generate 10,855 raw SCPR instances across 12 application domains, averaging 9.2 sample points and 6.8 reasons per instance.
  \item Discarded any duplicate SCPR instances, producing a final set of 2,851 unique SCPR files.
  \item Applied the generalized Beasley reduction to each unique instance, which (i) completely resolved 95.5\% of cases by trivial elimination, and (ii) shrunk the remaining instances by 63\% on average in terms of region–reason pairs.
\end{enumerate}

\begin{table}[ht]
  \centering
  \caption{Representative QF\_NRA benchmarks (SMT‐LIB 2024) and application domains}
  \label{tab:benchmarks}
  \begin{tabular}{ll}
    \toprule
    Benchmark directory               & Application domain \\
    \midrule
    UltimateAutomizer             & Program‐invariant synthesis \\
    zankl                         & Term‐rewriting termination \\
    meti‐tarski                   & Real‐algebraic theorem proving \\
    LassoRanker                   & Loop‐termination proof \\
    kissing                       & Discrete geometry \\
    hycomp                        & Hybrid‐systems reachability \\
    hong                          & Symbolic algebra \\
    pPDA‐Chiari‐Pontiggia‐Winkler & Analytic inequality \\
    Uncu                          & Inequality proving \\
    GeoGebra                      & Euclidean geometry \\
    Economics‐Mulligan            & Economic‐equilibrium modelling \\
    Heizmann‐UltimateInvariantSynthesis & Invariant synthesis \\
    \bottomrule
  \end{tabular}
\end{table}

\subsection{Results}

The experimental results revealed a decisive role for the reduction preprocessing and confirm that linear programming (LP) is the most consistently effective solver across all residual SCPR cases.

\begin{figure}[H]
\centering
\includegraphics[width=0.8\linewidth]{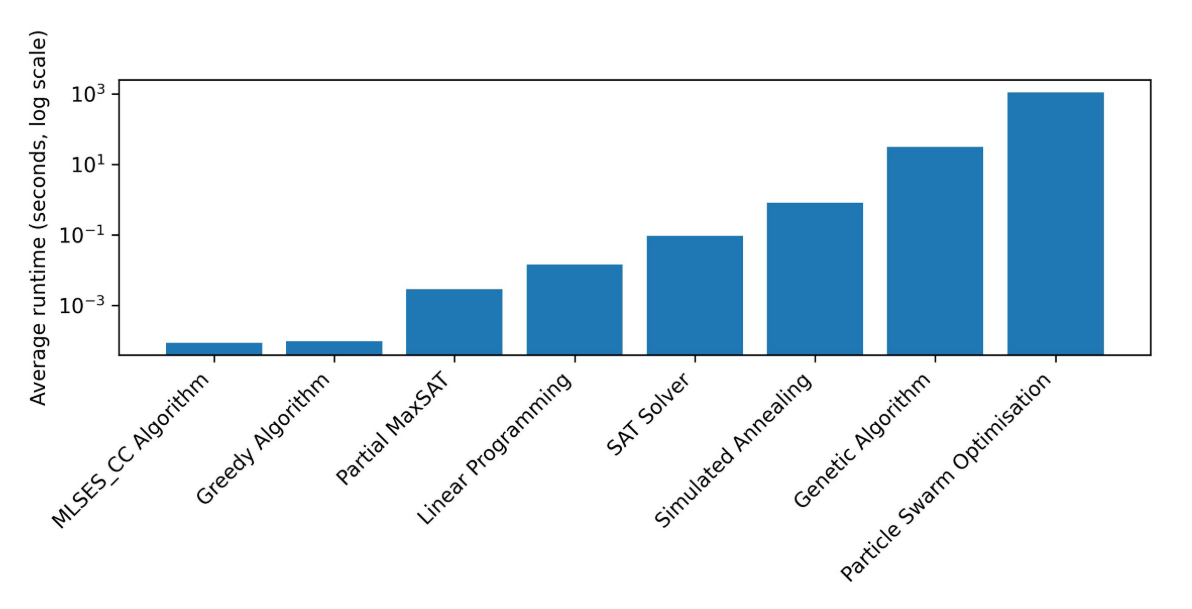}
\caption{Average run time per algorithm (logarithmic $y$-axis).}
\label{fig:runtime_global}
\end{figure}

Before introducing the Beasley reduction process, we evaluated eight solvers on 10,855 unreduced SCPR instances generated from CDCAC runs. Linear Programming (LP) serves as the benchmark, consistently producing optimal solutions against which the performance of the other algorithms is assessed.  Table~\ref{tab:global_perf} shows the global performance of all solvers in terms of average run time and accuracy, and Figure~\ref{fig:runtime_global} visualises these results.  Some key observations:

\begin{itemize}
    \item \textbf{LP and SAT} both achieved 100\% accuracy across all instances, as they should being exact methods.  LP was  faster on average ($0.095$ s vs $0.13$ s).
    \item \textbf{Partial MaxSAT} offered a good compromise: near-optimal accuracy ($98\%$) at a runtime of just $1.5 \times 10^{-3}$ seconds.
    \item \textbf{Greedy and MLSES-CC} were the fastest, finishing in microseconds, but sacrificed accuracy ($93\%$).
    \item \textbf{Metaheuristics (SA, GA, PSO)} showed good accuracy ($\approx$97\%), but with much higher runtimes; e.g. up to $1,080$ seconds for PSO.  I.e. they are slower than the exact methods, at least for these benchmarks.
\end{itemize}

\noindent
These results clearly establish LP as the most reliable baseline in terms of both correctness and practical feasibility. The results suggest that if problem instances require a trade-off of accuracy for speed then the partial MaxSAT approach is preferable to meta-heuristic methods: although we note the latter have received less intensive development and improvements may be possible.

\begin{table}[htbp!]
  \centering
  \caption{Performance statistics across 10,855 SCPR instances before Beasley reduction.}
  \label{tab:global_perf}
  \begin{tabular}{@{}lcc@{}}
    \toprule
    \textbf{Algorithm} & \textbf{Avg.\ run time [s]} & \textbf{Accuracy}\\
    \midrule
    MLSES-CC                     & $6.1\times10^{-5}$ & 0.93\\
    Greedy                        & $2.3\times10^{-4}$ & 0.93\\
    Partial MaxSAT                & $1.5\times10^{-3}$ & 0.98\\
    Linear Programming (baseline) & $9.5\times10^{-2}$ & 1.00\\
    SAT re-encoding               & $1.3\times10^{-1}$ & 1.00\\
    Simulated Annealing           & $4.9$              & 0.97\\
    Genetic Algorithm             & $29$               & 0.97\\
    Particle Swarm Optimisation   & $1.1\times10^{3}$  & 0.89\\
    \bottomrule
  \end{tabular}
\end{table}

The performance statistics reported in Table~\ref{tab:global_perf} and the runtime results reported in Figure~\ref{fig:runtime_global}, however, are computed over the full set of $10 855$ benchmark instances, including duplicates. This choice reflects realistic SMT solver usage, where identical or structurally equivalent SCPR problem instances frequently arise across different solver states and incremental calls. Retaining duplicates captures the cumulative cost incurred when reductions and solving procedures are repeatedly applied during CDCAC execution. Importantly, including duplicates does not distort the qualitative conclusions: since the Beasley reduction deterministically eliminates the same instances regardless of repetition.

After removing duplicate inputs, we obtained 2\,851 unique SCPR instances derived from the CDCAC benchmark runs. Applying the Beasley reduction procedure immediately solved 2\,722 of these instances: a 95.5\% success rate without any substantial algorithmic search. Only 129 instances (4.5\%) required further solving.  This represents a significant insight: for common CDCAC workloads, the Beasley reduction process serves as a near-complete decision procedure. Solvers not interested in spending considerable effort on this optimisation could still perform at least this reduction and reap most of the benefit.

For the remaining 129 instances, our binary LP formulation solved every case optimally in under 1 millisecond. No other algorithm (heuristic or exact) achieved better performance on these.  
This suggest that together, the reduction and LP form a lightweight, robust pipeline for conflict core minimisation within CDCAC-based SMT solvers.

Table~\ref{tab:beasley-lp-per-domain} breaks down the performance of the Beasley + LP combination by application domain. Even in domains where the reduction step was less effective, such as real-algebra proving (QF\_NRA\_3) and inequality proving (QF\_NRA\_9), the LP solver maintained perfect accuracy with negligible overhead.

\begin{table}[ht]
  \centering
  \caption{Beasley reduction and LP solving per QF\_NRA application domain.}
  \label{tab:beasley-lp-per-domain}
  \begin{tabular}{lrrr}
    \toprule
    Domain (directory) & Unique & Beasley‐Solved & LP‐Solved \\
    \midrule
    1 (Invariant synthesis)         &  400 & 368 (92.0\%) &  32 (8.0\%) \\
    2 (Termination analysis)        &  169 & 162 (95.8\%) &   7 (4.2\%) \\
    3 (Real‐algebra proving)        &   37 &  16 (43.2\%) &  21 (56.8\%) \\
    4 (Loop‐termination proof)      & 1547 & 1521 (98.3\%)&  26 (1.7\%) \\
    5 (Discrete geometry)           &   37 &  29 (78.4\%) &   8 (21.6\%) \\
    7 (Symbolic algebra)            &    4 &   4 (100\%)  &   0 (0\%)   \\
    8 (Analytic inequality)         &   30 &  26 (86.7\%) &   4 (13.3\%)\\
    9 (Inequality proving)          &   31 &  16 (51.6\%) &  15 (48.4\%)\\
    10 (Euclidean geometry)         &    8 &   7 (87.5\%) &   1 (12.5\%)\\
    12 (Economic modelling)         &   72 &  71 (98.6\%) &   1 (1.4\%) \\
    13 (Invariant synthesis)        &  513 & 502 (97.8\%) &  11 (2.2\%) \\
    14 (Hybrid‐systems reachability)&    3 &   0 (0\%)    &   3 (100\%) \\
    \midrule
    \textbf{Total}                  &2,851 &2,722 (95.5\%)& 129 (4.5\%) \\ 
    \bottomrule
  \end{tabular}
\end{table}

\newpage 

\section{Conclusions and Future Work}
\label{sec:Conc}

In this paper, we have identified and formalised the SCPR as the key optimisation task underlying conflict generalisation when using CDCAC SMT solving in non‐linear real arithmetic.  We demonstrated that, in practice, SCPR instances arising from CDCAC can be efficiently handled by a two‐phase pipeline of reduction and linear programming.

Our main contributions are:
\begin{itemize}
  \item Problem formalisation: We extended the classical Set Covering Problem to SCPR, capturing multi‐constraint conflicts in higher‐dimension CAD cells.
  \item Generalised Beasley reduction: We adapted Beasley’s data‐reduction rules~\cite{BEASLEY1987851} to SCPR, enabling elimination of 95.5\% of CDCAC‐derived instances purely syntactically.
  \item Solver evaluation: For the 4.5\% of instances remaining after reduction, we showed that a linear‐programming (LP) formulation finds optimal reason sets in under 1\,ms per instance, outperforming SAT-based and heuristic alternatives.
\end{itemize}

These findings establish the SCPR as a practical framework for clause minimisation in CAD‐based SMT.  Integrating this pipeline into CDCAC would yield provably minimal conflict clauses with almost zero overhead.

\paragraph{Future work} Building on these results, we see several promising directions:
\begin{enumerate}
  \item \textbf{In‐solver integration:} Embed the generalised Beasley reduction and LP–based solver directly into CDCAC engines (e.g., SMT‐RAT, cvc5, Maple) for \emph{on‐the‐fly} clause minimisation.
  \item \textbf{Hard‐case analysis:} Investigate specialised reduction and solving strategies for the challenging instances in the geometric and inequality-proving domains, where reduction yields limited results.
\end{enumerate}

\begin{acknowledgments}
The first author was supported by a scholarship from Coventry University.  The second and third authors were supported by UKRI EPSRC grant EP/T015748/1: ``\emph{Pushing Back the Doubly-Exponential Wall of Cylindrical Algebraic Decomposition}'' (the DEWCAD Project).
\end{acknowledgments}

\section*{Declaration of Competing Interest}

The authors declare that they have no known competing financial interests or personal relationships that could have appeared to influence the work reported in this paper.

\section*{Data Access Statement}

The sources QF\_NRA data and the resulting SCPR problem instances are available via Zenodo: \href{https://doi.org/10.5281/zenodo.15326494}{https://doi.org/10.5281/zenodo.15326494}

\bibliographystyle{elsarticle-num}
\bibliography{references_ejor}

\begin{comment}

\end{comment}

\end{document}